# A durable and efficient electrocatalyst for saline water splitting with current density exceeding 2000 mA cm$^{-2}$

Fengning Yang[1†], Yuting Luo[1†], Qiangmin Yu[1], Zhiyuan Zhang[1], Shuo Zhang[2], Zhibo Liu[3], Wencai Ren[3], Hui-Ming Cheng[1,3], Jiong Li[2*] & Bilu Liu[1*]

1. Shenzhen Geim Graphene Center, Tsinghua-Berkeley Shenzhen Institute & Tsinghua Shenzhen International Graduate School, Tsinghua University, Shenzhen 518055, P. R. China.

2. Shanghai Synchrotron Radiation Facility, Shanghai Advanced Research Institute, Chinese Academy of Sciences, Shanghai 201210, P. R. China.

3. Shenyang National Laboratory for Materials Sciences, Institute of Metal Research, Chinese Academy of Sciences, Shenyang, Liaoning, 110016, P. R. China.

† These two authors contributed equally.

* E-mail: B.L. (bilu.liu@sz.tsinghua.edu.cn) or J.L. (lijiong@sinap.ac.cn)

**ABSTRACT**

Water electrolysis is promising for industrial hydrogen production to achieve a sustainable and green hydrogen economy, but the high cost of the technology limits its market share. Developing efficient yet economic electrocatalysts is crucial to decrease the cost of electricity and electrolytic cell. Meanwhile, electrolysis in seawater electrolyte can further reduce feedstock cost. Here we synthesize a type of electrocatalyst where trace precious metals are strongly anchored on corrosion-resistive matrix. As an example, the produced Pt/Ni-Mo electrocatalyst only needs an overpotential of 113 mV to reach an ultrahigh current density of 2000 mA cm$^{-2}$ in saline-alkaline electrolyte, standing as the best performance so far. It shows high activity and long durability in various electrolytes and under harsh conditions, including strong alkaline and simulated seawater electrolytes, and under elevated temperatures up to 80 ºC. This electrocatalyst is produced on a large scale at low cost and shows good performance in a commercial membrane electrode assembly stack, demonstrating its feasibility for practical water electrolysis.

Environmental problems and the energy crisis caused by the excessive use of fossil fuels have raised a need for renewable energy technologies. Hydrogen which has the highest mass energy density (three times than that of gasoline) is a promising energy source. Water electrolysis is expected to be widely used for industrial hydrogen production to achieve a sustainable and green hydrogen economy, especially when it is powered by electricity from renewable energy such as sunlight or wind, together with the use of low-grade water or seawater.[1] However, water electrolysis technology is still far from being economically competitive with other technologies, such as steam methane reforming and coal gasification, mainly because the sluggish kinetics of the hydrogen evolution (HER) and oxygen evolution (OER) half reactions contributed to the high electricity consumption.[2] To solve this issue, platinum group metal (PGM)-based electrocatalysts with a high intrinsic activity have been used, but their high cost and scarcity greatly increase the catalyst price, which accounts for ~8% of the stack cost of electrolyzers.[3] The development of cheap yet efficient catalysts for water splitting is a priority. Loading small amount of PGM on cheap supports is a strategy to reduce the catalyst price while maintaining high activity. Many efforts to do this have been made in recent years, including the loading of Pt single atoms on supports, low-Pt multicomponent catalysts, bimetallic Pt alloy catalysts, *etc*.[4] For example, Zhang *et al*. reported that Pt single atoms anchored on the MXene nanosheets enhanced the HER performance, showing an overpotential of 30 mV at 10 mA cm$^{-2}$.[4c] Xing *et al*. synthesized a catalyst made of ultrafine Pt nanoparticles loaded on Co(OH)$_2$ nanosheets that showed an overpotential of 185 mV at 200 mA cm$^{-2}$ for HER.[5] By using a partial electrochemical dealloying method, Li *et al*. prepared Pt nanowires modified by nickel single atoms that had both a high specific activity and an electrochemical surface area (ECSA) for HER and other reactions.[6] Despite these achievements, electrocatalysts that perform well at high currents and face to practical applications are still rarely reported.

Besides economic considerations, catalysts with high efficiency and durability at high current densities are required for practical applications. A commercial advanced alkaline water electrolyzer usually needs to operate at current densities in the range from 200 to 1000 mA cm$^{-2}$, and even up to 2000 mA cm$^{-2}$ in an anion exchange membrane (AEM) system, combined with a high alkali concentration (~6 M KOH) and an elevated temperature (70–90 °C) to reduce the overpotentials.[7] Raney nickel is a commercial electrocatalyst used in alkaline water electrolyzers, but it needs a large overpotential for HER to reach a high current density, *e.g.*, ~200 mV (1 M KOH, 25 °C) and ~100 mV (6 M KOH, 80 °C) at 500 mA cm$^{-2}$ for the best Raney Ni catalyst reported so far.[8] Good performance at a high current density is a crucial challenge in electrocatalysis due to the complex behavior in the surface of catalysts under the influence of violent current and airflow. A few researchers have started to design new materials to overcome this challenge by modulating the geometrical structure, the size of active sites, or catalyst compositions.[9] For instance, Ge *et al.* reported that using a three-dimensional Prussian blue analogue support, Ni$_2$P/Fe$_2$P electrocatalyst delivered 500 mA cm$^{-2}$ at an overpotential of 226 mV for HER.[9a] Our group recently reported that by engineering morphology and surface chemistry, a MoS$_2$/Mo$_2$C catalyst showed high performance for HER with an overpotential of 220 mV at 1000 mA cm$^{-2}$.[10] In addition to the efficiency at a high current density, the durability of the catalyst is another factor in evaluating the performance. This requirement is not only focused on good chemical and mechanical stabilities, but also hope to achieve the long-term stability in a variety of extreme and practical conditions, such as a strong alkali, saline water, or even seawater and other low-quality water, which can reduce the burden of freshwater.[11] In this regard, nickel-molybdenum (Ni-Mo) alloys or compounds with superior corrosion resistance have been widely studied in seawater splitting.[12] For example, Luo *et al*. prepared an efficient NiMoN catalyst, where an overpotential of 170 mV was needed to reach a current density of 1000 mA cm$^{-2}$ in simulated seawater solution.[13]

Here, we report the development of an electrocatalyst where pre-embedded PGM is controllably reduced from a corrosion-resistant Ni-Mo matrix by a reduction-potential dependent sequential reduction process, resulting in highly dispersed PGM nanoparticles strongly anchored on the matrix. Taking examples, Pt/Ni-Mo catalyst shows an overpotential of 42 mV for HER and Ru/Ni-Mo catalyst shows an overpotential of 420 mV for OER at the high current density of 2000 mA cm$^{-2}$ in 1 M KOH solution. The highly dispersed PGM nanoparticles and the pre-designed matrix contribute to the extraordinary activity of the catalysts. Impressively, Pt/Ni-Mo catalyst operates well in saline-alkaline water (1 M KOH and 0.5 M NaCl) with an overpotential of 113 mV at 2000 mA cm$^{-2}$ and keeps long durability in various electrolytes under harsh conditions including a strong alkaline electrolyte or a saline-alkaline electrolyte (24 hours), an ultrahigh current density of 2000 mA cm$^{-2}$ (140 hours), and elevated temperatures up to 80 ºC. We also show the feasibility of the scale-up production of such an electrocatalyst and its use in a commercial membrane electrode assembly (MEA) stack. Noteworthy, this kind of catalysts provide excellent performance in sustainable applications and essentially improve the economic competitiveness of green hydrogen production by electrolysis.

We proposed a sequential reduction strategy for synthesizing high-performance catalysts, where PGMs are pre-embedded in a matrix first and then preferentially occur reductive reaction by hydrogen due to their low reduction potentials and immobilized on its surface (Figure 1a). The catalysts are synthesized by this strategy with a low amount of PGM that is highly dispersed and strongly anchored on a large-surface-area and corrosion-resistive matrix. As a proof of concept, a trace amount of Pt or Ru was mixed with nickel and molybdate precursors, followed by coprecipitation and hydrogen reduction to prepare Pt/Ni-Mo and Ru/Ni-Mo catalysts. During hydrogen reduction, PGMs preferentially occurred reductive reaction because their oxide has lower reduction potentials than the matrix materials (Supplementary Figure 1). Specifically, the standard molar enthalpies of formation $\Delta H^{\ominus}$ (298.15 K) of $PtO_2$ and $RuO_{3/2}$ are −167 ( ± 42) kJ mol$^{-1}$ and −130 ( ± 20) kJ mol$^{-1}$, while those of $MoO_3$ and NiO are lower than −200 kJ mol$^{-1}$.[14] X-ray diffraction (XRD) and scanning electron microscopic (SEM) results of Pt/Ni-Mo before and after hydrogen reduction were compared and indicated that adding Pt did not change the crystal structure of the matrix but reduced the extent of reduction of the matrix. (Supplementary Figures 2 and 3). These results confirm that the design principle of such a sequential reduction strategy is feasible.

We then conducted the electrochemical tests of Pt/Ni-Mo and Ru/Ni-Mo catalysts for water splitting in 1 M KOH solution for evaluating their activities. For the HER, Pt/Ni-Mo need overpotentials of 33 mV and 42 mV at current densities of 1000 mA cm$^{-2}$ and 2000 mA cm$^{-2}$, respectively, which are over 10 times lower than that of Ni foam loaded with 2 mg cm$^{-2}$ 20% Pt/C and over 20 times lower than that of a pure Pt foil at 2000 mA cm$^{-2}$ (Figure 1b). The Pt/Ni-Mo catalyst with a 0.76 wt% Pt loading shows the best HER performance among all PGM-based and non-PGM catalysts reported so far (Figure 1c and Supplementary Table 1). For the OER, Ru/Ni-Mo needs overpotentials of 390 mV and 420 mV to reach current densities of 1000 mA cm$^{-2}$ and 2000 mA cm$^{-2}$, respectively which are almost three times lower than that of Ni foam loaded with 3 mg cm$^{-2}$ $IrO_2$ (Figure 1d and Supplementary Table 2). Similar to Pt/Ni-Mo for HER, the Ru/Ni-Mo also shows excellent performance for OER (Figure 1e). Overall, these results show that electrocatalysts synthesized by sequential reduction strategy have a record high performance in alkaline media for

HER and OER.

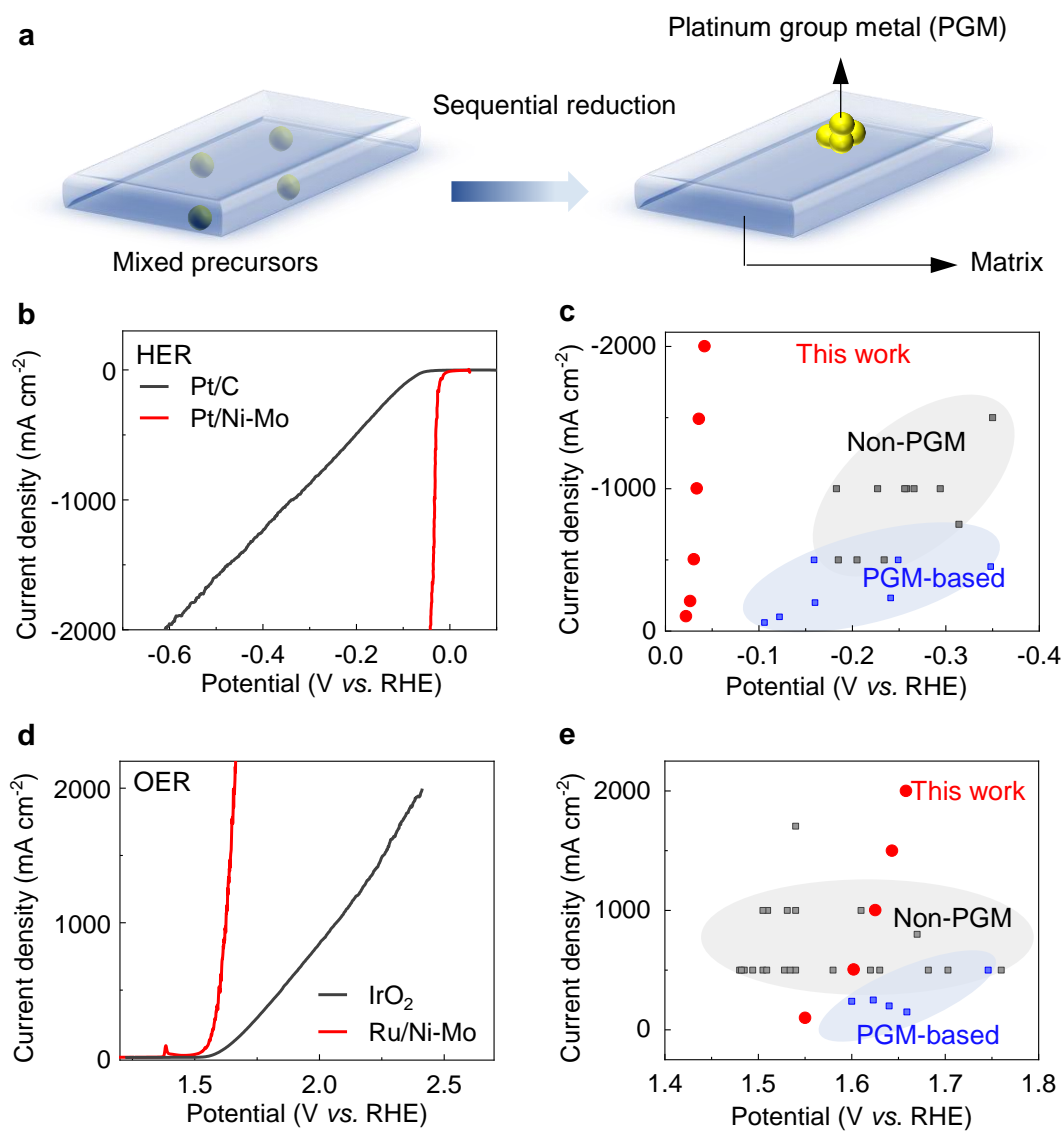

**Figure 1. Design principle of the sequential reduction strategy and the high-performance of such catalysts in 1 M KOH solutions. a** Reduction potential-dependent sequential reduction of PGMs in matrix. **b, d** Polarization curves of (b) Pt/Ni-Mo electrocatalysts for HER and (d) Ru/Ni-Mo catalysts for OER in 1 M KOH solutions. These data are with *iR* compensations (85%). **d, g** Comparison of the activities of these catalysts with state-of-the-art (d) HER and (g) OER electrocatalysts in 1 M KOH solutions.

Taking Pt/Ni-Mo catalysts as an example, we studied its synthesis and structure in detail. As illustrated in Figure 2a, the Pt/Ni-Mo was synthesized by two steps. First, $H_2PtCl_6$, $(NH_4)_2MoO_4·7H_2O$, and $Ni(NO_3)_2·6H_2O$ salts were dissolved in an aqueous solution to grow the $Pt/NiMoO_4$ on Ni foam by a hydrothermal process. Second, $Pt/NiMoO_4$ was reduced by hydrogen to prepare Pt/Ni-Mo catalyst (see details in the "Methods" section). In the first step, $Pt/NiMoO_4$ micro-columns were grown on the Ni foam. After the hydrogen reduction, Pt nanoparticles and $Ni_4Mo$ nanocrystals were sequential released and anchored on the surface of $NiMoO_4$ matrix with the micro-column morphology maintained well, following the reaction,

$$Pt/NiMoO_4·H_2O \xrightarrow{H_2,\ 500\ °C} Pt + Ni_4Mo + NiMoO_4(Matrix) \quad (1)$$

We synthesized materials using different parameters to optimize the HER performance of the catalysts (Supplementary Figures 4−9). Noted that the roughness of column arrays is modulated by adjusting temperatures and time for hydrogen reduction (Supplementary Figures 6−9 and Supplementary Table 3). With the increasement of hydrogen reduction temperature, the diameters of particles on the $NiMoO_4$ columns increased from 30 nm to 50 nm (Supplementary Figures 6 and 7). The density of particles on columns increases with increasing reduction time, and the particles are seriously aggregated after reduction for 3 hours (Supplementary Figures 8 and 9). The sample with the best HER performance was obtained when the reduction was conducted at 500 °C for 15 min.

We also carried out structural characterization of the optimized Pt/Ni-Mo catalyst. SEM images show that dense micro-column arrays with a length of tens of microns and width of 0.5 to 1.0 millimeter are grown on the nickel foam. Many nanoparticles with sizes of 30 to 40 nm are uniformly dispersed over the surface of the columns (Figure 2b and Supplementary Figure 1). Elemental mapping of energy dispersive X-ray spectroscopy (EDS) and inductively coupled plasma optical emission spectroscopy (ICP-OES) confirm that the Pt/Ni-Mo columns consist of Pt, Ni, Mo and O elements (Figure 2c and Supplementary Table 4). EDS results indicate that the atomic ratio of Ni/Mo is 0.85 in these columns (Supplementary Figure 10 and Supplementary Table 5). ICP-OES results show that the amount of Pt is 0.76 wt%. The XRD results (Figure 2d) show that the Pt/Ni-Mo contains $NiMoO_4$ (JCPDS, No. 33-0948), $MoO_2$ (JCPDS, No. 65-5480), and $Ni_4Mo$ (JCPDS, No. 32-0671). High-resolution transmission electron microscopy (HRTEM) shows that different components are found in the inside and the outside regions of the catalyst (Figures 2e and 2f). In the inside region, the lattice fringe with an interplanar spacing of 0.31 nm is attributed to the $NiMoO_4$ (220) planes (Figure $2f_1$). In the outside region (edges of the sample), domains with a lattice spacing of 0.22 nm correspond to Pt (111) planes. Besides Pt, we also find $Ni_4Mo$ (002) and $MoO_2$ (211) planes (Figure $2f_2$ and $2f_3$) and these correspond to spots in the selected area electron diffraction (SAED) pattern (Figure 2g). The X-ray photoelectron spectroscope (XPS) results show peaks of $Mo^0$ and $Ni^0$ at 852.5 eV and 228.3 eV (Figures 2h and 2i), confirming the existence of metallic Mo and Ni in the surfaces of Pt/Ni-Mo columns. Because of the extremely low Pt content, we used X-ray absorption spectroscopy (XAS) to study its state (Supplementary Figure 11). Both the X-ray absorption near edge structure (XANES) (Figure 2i) and the oscillation in the Fourier transformed extended X-ray absorption fine structure (EXAFS) of Pt $L_3$-edge (Figure 2j) of Pt/Ni-Mo are similar to these of Pt foil, indicating that the Pt in Pt/Ni-Mo catalyst is in the metallic state. The decrease in Pt−Pt bond length and oscillation intensity suggest that the sizes of Pt particles are small and they are well dispersed on matrix (Supplementary Figure 12).[15] These results show that the Pt/Ni-Mo is composed of $NiMoO_4$ column matrix with Pt nanoparticles as well as $Ni_4Mo$ and $MoO_2$ domains anchored on its outer surface as illustrated in Figure 2k.

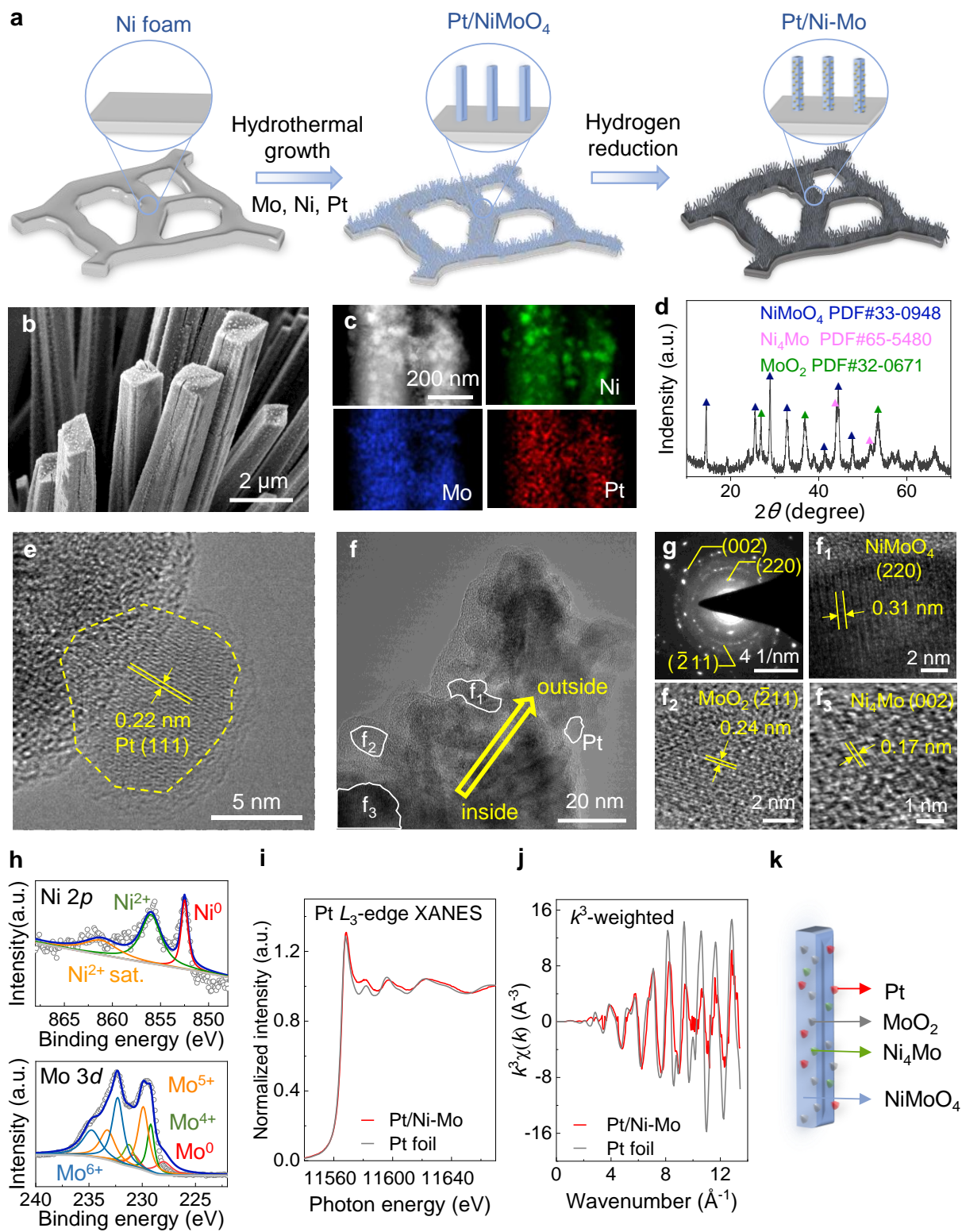

**Figure 2. Synthesis and characterization of the Pt/Ni-Mo catalyst. a** Schematic of the two-step synthesis process. **b** A SEM image of Pt/Ni-Mo. **c** A STEM image and corresponding EDS elemental maps showing the distribution of Ni, Mo, and Pt. **d** A XRD pattern. **e** HRTEM image of Pt nanoparticles

in Pt/Ni-Mo. **f** Magnified HRTEM images of Pt/Ni-Mo. **g** SAED pattern of the area in f. **f₁-f₃** Lattice fringes seen areas in f. **h** XPS spectra of Ni 2*p* (top) and Mo 3*d* (down). **i** XANES of the Pt $L_3$-edge. **j** The $k^3$-weighted Pt $L_3$-edge EXAFS spectra. **k** Schematic of the structure of Pt/Ni-Mo catalyst.

To shine some lights on the excellent HER performance of Pt/Ni-Mo catalyst, its mass transfer and charge transfer abilities were studied. The low interfacial adhesion energy of the catalyst can directly improve the transfer of products and reactants, especially at high current densities. We first analyzed the wettability of Pt/Ni-Mo in terms of static contact angles ($\theta$) and dynamic contact hysteresis ($\theta_h$) and compare it with three reference materials including carbon paper, Ni foam, and Pt foil. The static contact angle results show that the Pt/Ni-Mo is superhydrophilic, with a $\theta$ close to zero (Figure 3a), while the other three catalysts show much larger $\theta$ (in the range of 60°–120°) and therefore poorer affinities to electrolyte droplets. We also measured the dynamic contact hysteresis $\theta_h$ by the Wilhelmy balance method to mimic catalytic electrodes immersed in electrolytes, by subtracting the receding ($\theta_r$) from the advancing ($\theta_a$) contact angles as follows,[16]

$$\theta_h = \theta_a - \theta_r \qquad (3)$$

The small $\theta_h$ of Pt/Ni-Mo (3°) indicates a low interfacial adhesion which may result in the easy mass transfer of electrolytes and removal of hydrogen bubbles on its surface (Figures 3b and 3c). Collectively, these results show a good mass transfer ability on the Pt/Ni-Mo catalyst.[10, 17]

Besides mass transfer ability, the large number of highly dispersed Pt nanoparticles is another reason for the good HER performance of Pt/Ni-Mo catalysts. We measured its ECSA by the electrochemical double-layer capacitance ($C_{dl}$) method and the specific surface area by Brunauer-Emmett-Teller (BET) method (Supplementary Figures 13-15). Pt/Ni-Mo has a considerably larger specific surface area (50.51 m$^2$ g$^{-1}$) and electrochemical surface area (582.2 cm$^2$) per square centimeter electrode than the other three catalysts (Figure 3d), suggesting a large available surface area to support highly-dispersed Pt nanoparticles on it. We then compared the specific activity of the catalyst normalized by its ECSA with several PGM-based catalysts in previous reports (Supplementary Table 6), and found that its specific current density is higher than others (Figure 3e). These results show that it has a high intrinsic activity and exposes abundant active sites to the electrolyte. We also compared the $\Delta\eta/\Delta\log|j|$ ratio of these catalysts at different current densities, which means how much overpotential is needed when the current increases and is a good indicator of the performance of the catalyst over a broad current density range.[10, 18] The $\Delta\eta/\Delta\log|j|$ ratio of Pt/C increases sharply when the current density is larger than 200 mA cm$^{-2}$, and reaches 1203 mV dec$^{-1}$ at 2000 mA cm$^{-2}$ (Figure 3f). In sharp contrast, the ratio for Pt/Ni-Mo catalyst remains small, only 27 mV dec$^{-1}$ at 2000 mA cm$^{-2}$, indicating its excellent HER performance at high current densities. We find that the Pt/Ni-Mo catalyst shows the highest current densities at overpotentials of 10, 30 and 40 mV among all catalysts, including precursor Pt/NiMoO$_4$, support Ni foam, Pt/C and the Ni-Mo without Pt (Figure 3g), which can be attributed to the pre-designed matrix and high intrinsic activity of Pt. These results show that the Pt/Ni-Mo catalyst has a good performance for high-current-density HER due to its combined effects of good mass transfer ability and large numbers of accessible sites with high intrinsic activity.

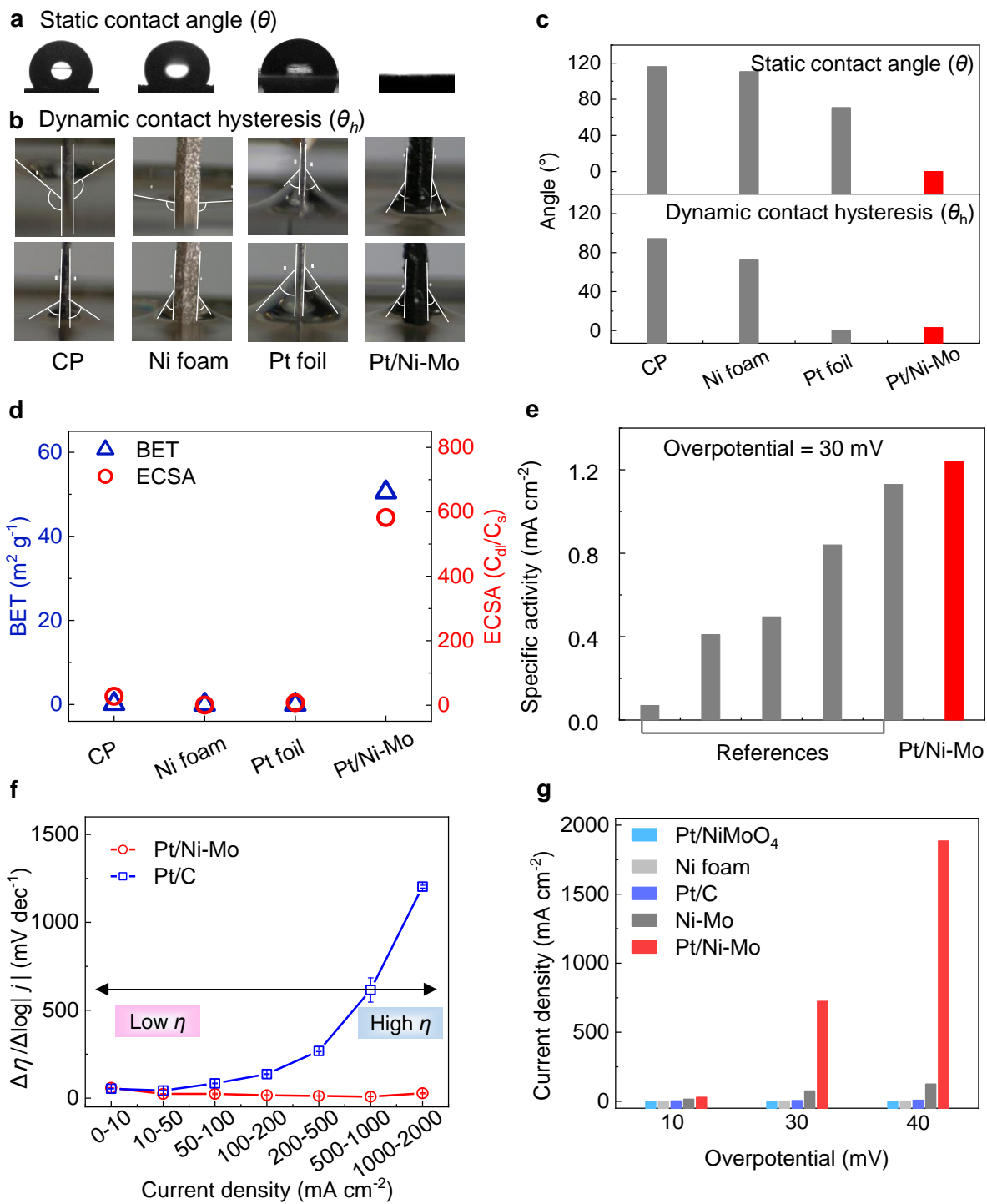

**Figure 3. Study of the mechanism HER performance of Pt/Ni-Mo catalyst. a** Optical images showing the static contact angles ($\theta$) and **b** dynamic contact angle hysteresis ($\theta_h$) of electrolyte drops on different catalysts, including carbon paper, Ni foam, Pt foil, and Pt/Ni-Mo. **c** Histogram of contact angles. **d** BET surface area and ECSA. The ECSA is determined as the ratio of $C_{dl}/C_s$, where $C_{dl}$ is the measured capacitance and $C_s$ is the specific capacitance of the catalysts. **e** Comparison of the specific activities normalized by ECSA for PGM-based electrocatalysts. **f** The $\Delta\eta/\Delta\log|j|$ ratio for the Pt/C and Pt/Ni-Mo in different current density ranges. **g** Current densities at overpotentials of 10, 30 and 40 mV for different catalysts in 1 M KOH solutions.

We tested the performance and the long durability of the catalyst in various electrolytes under the harsh conditions. A three-electrode system was used and the distance between the working electrodes and reference electrode were optimized and remained the same in each run to avoid the influence of solution resistance (Supplementary Figure 16). We first compared the HER performance of Ni foam, Pt foil, and Pt/Ni-Mo in 1 M and 6 M KOH solutions, and found that Pt/Ni-Mo has the best performance (Figure 4a and Supplementary Figure 17). Then, a mixture of 0.5 M NaCl and 1 M KOH was used as the simulated seawater electrolyte (*i.e.*, saline-alkaline electrolyte), in which the Pt/Ni-Mo needed a small overpotential of 113 mV to reach 2000 mA cm$^{-2}$, the best performance reported so far in literature (Figure 4b and Supplementary Table 7). The Faradaic efficiency of Pt/Ni-Mo for hydrogen production in 1 M KOH and saline-alkaline solutions were both measured to be ~100% (Figure 4c and Supplementary Table 8). These results suggest that Pt/Ni-Mo catalyst has a high efficiency in different electrolytes.

We then focused on the electrochemical durability of the catalyst. A H-type electrolytic cell was used for all long-term tests so that the influence of oxygen on the counter electrode could be avoided. The results showed a negligible increase of the potential after 74 hours in 1 M KOH solution as the current density increased from 300 to 500 and further to 1000 mA cm$^{-2}$ (Supplementary Figure 18). Moreover, it showed almost no increase in potential after operating for 140 hours at 2000 mA cm$^{-2}$ (Figure 4d), suggesting the excellent long-time durability of Pt/Ni-Mo catalyst in a 1 M KOH electrolyte at an ultrahigh current density. In addition to that, the catalyst was evaluated in even practical conditions. For example, we found that the catalyst operated well in 6 M KOH solution for over 24 hours, and even under high temperature conditions at 25, 40, 50, 60 and 80 °C (at a current density of 500 mA cm$^{-2}$) (Figures 4e and 4f). we also tested the durability of the catalyst at 500 mA cm$^{-2}$ in saline alkaline solution composed of 1 M KOH and 0.5 M NaCl, because chloride corrosion is a big challenge to the catalyst durability.[19] Note that the electrolysis remained stable for more than 24 h, without obvious corrosion or a voltage increase, suggesting its good stability in the saline water electrolysis. These results not only indicate the superior durability of our catalyst at high current densities but also demonstrate its feasibility in different practical applications, especially for seawater splitting.

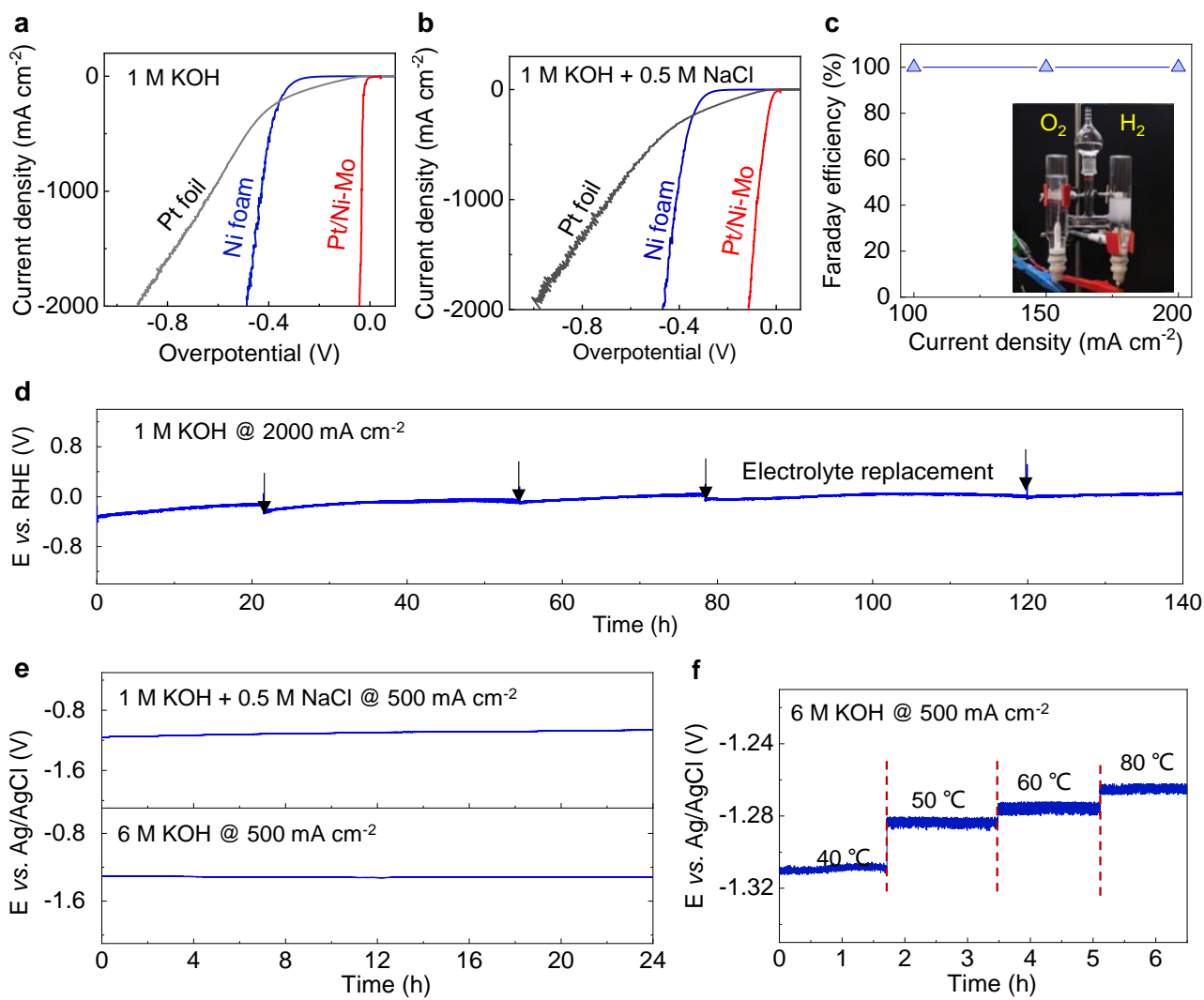

**Figure 4. HER performance and durability of Pt/Ni-Mo catalyst at different practical conditions.**
**a, b** Polarization curves with *iR* compensation for Ni foam, Pt foil and Pt/Ni-Mo in the 1 M KOH (a) and 1 M KOH and 0.5 M NaCl mixed solution (b). **c** Faradaic efficiency of HER on Pt/Ni-Mo at different current densities of 100, 150 and 200 mA cm$^{-2}$ in saline alkaline solution. The insert is a photograph of the measurement device. **d** Chronopotentiometry (CP) curve for HER using Pt/Ni-Mo at 2000 mA cm$^{-2}$ for a total of 140 h. The electrolyte was a 1 M KOH aqueous solution. **e** CP curves for HER using Pt/Ni-Mo in 1 M KOH + 0.5 M NaCl and 6 M KOH aqueous solutions at 500 mA cm$^{-2}$ for 24 hours. **f** CP curves for HER using Pt/Ni-Mo at 40, 50, 60 and 80 °C at 500 mA cm$^{-2}$. The electrolyte was a 6 M KOH aqueous solution.

For industrial applications, a catalyst electrode in a cylindrical vessel of alkaline water-electrolysis electrolyzer is usually larger than 500 cm$^2$, and even reaches several square meters.[7] We prepared a Pt/Ni-Mo catalyst with an area of 700 cm$^2$ (10 cm in width and 70 cm in length, Figure 5a) to show the scaling up production capability. The catalyst has a uniform appearance over the whole substrate and an identical morphology to the one with small dimensions (Figures 5b and 2b). The Pt/Ni-Mo catalyst was used in a MEA electrolyzer of a commercial hydrogen generator (Figure 5c). One MEA stack consisted of two HER cells on the sides and two OER cells in the middle (Figure 5d). The alkaline exchange membrane separates the HER cells and OER cells to avoid gas diffusion. The system uses Pt/Ni-Mo as the cathode and nickel foam as the anode, and requires a lower voltage than the original commercial hydrogen generator which uses Pt and Ir/Ru catalysts (*e.g.*, 2.06 V *vs* 2.63 V to reach 600 mA) in alkaline solutions (Figure 5e). Taking all things into account, the self-supported Pt/Ni-Mo catalyst shows its potential to be directly used in a commercial MEA stack.

To highlight the advantages of the Pt/Ni-Mo electrocatalyst, we compared its price and high-current-density performance with other reported catalysts (Figure 5f and Supplementary Tables 9−11). Overall, PGM-based catalysts have obvious advantages in performance, but their cost increases the capital cost. While non-PGM catalysts are cheaper but usually exhibit much poorer performance than PGM especially at large current densities (Supplementary Figure 19 and Table 12). Note the Pt/Ni-Mo has a price of ~US$ 320 per square meter, much cheaper than other PGM-based catalysts and comparable with cheap non-PGM catalysts. Meanwhile, Pt/Ni-Mo has the best performance ($\eta_{1000}$ is 33 mV for HER) at a high current density, maximizing the high activity advantages of noble metals. Therefore, the catalyst is economically competitive as considering its power-hydrogen conversion efficiency and stack capital cost (Figure 5f). The strategy presented here is universal and may guide

the rational design of catalysts using other precious metals (such as Pd, Ir) for different electrochemical technologies (Supplementary Figure 20).

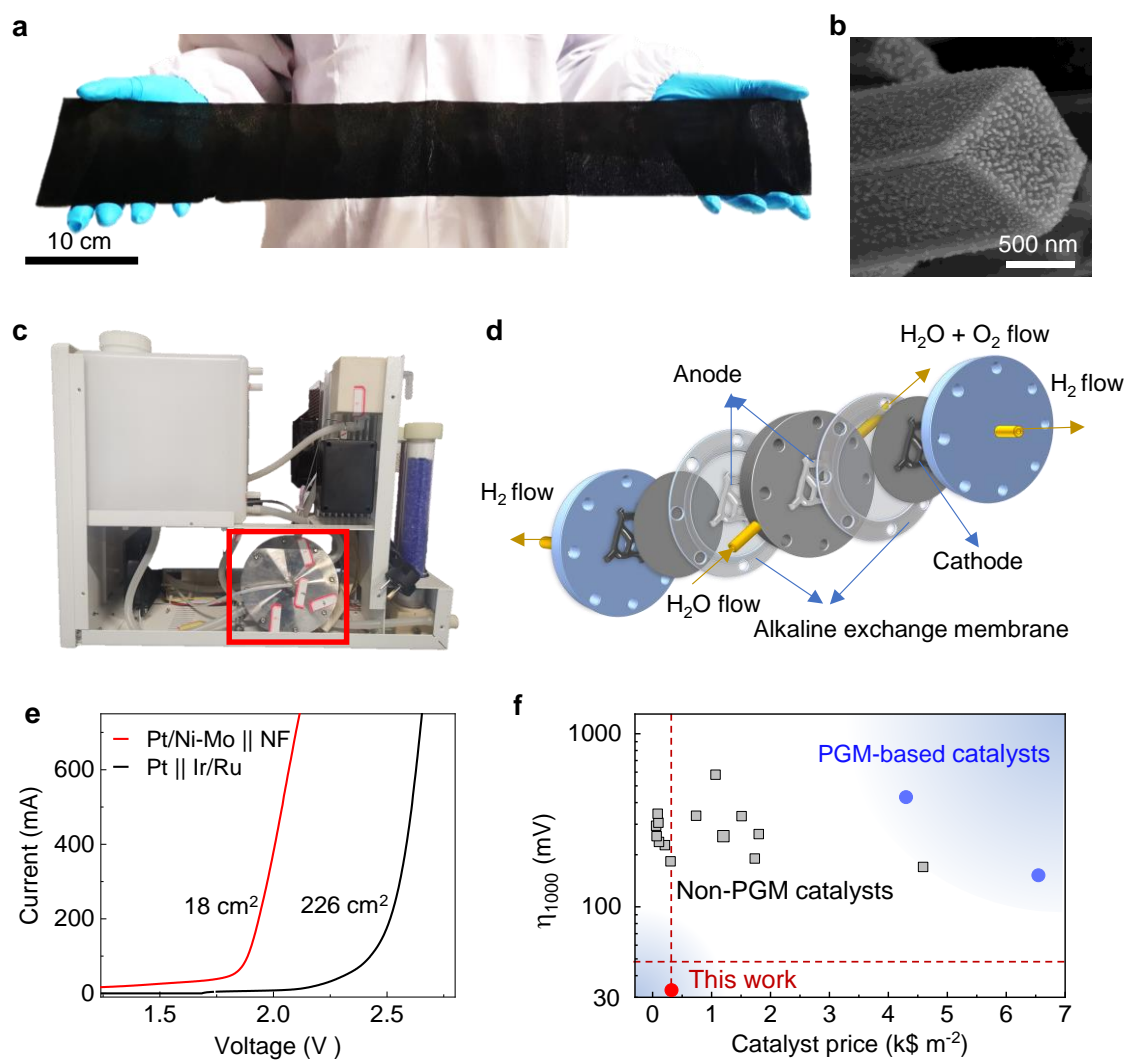

**Figure 5. Scaled-up production of Pt/Ni-Mo catalyst and its use in a commercial hydrogen generator. a** Pt/Ni-Mo catalyst with an area of 700 cm$^2$ (10 cm in width and 70 cm in length). **b** A SEM image of the Pt/Ni-Mo catalyst synthesized on a large scale. **c** A photo of the TH-1000 Hydrogen Generator (Beijing BCHP Analytical Technology Institute, China). **d** Stack structure of the electrolysis cell used in the machine testing. **e** Polarization curves in an electrolysis cell for the original membrane electrode of the TH-1000 Hydrogen Generator and Pt/Ni-Mo on Ni foam (NF) in a 1 M KOH solution. **f** Comparison of the price and large-current performance of Pt/Ni-Mo catalyst and other catalysts.

In conclusion, we have developed a sequential reduction strategy to synthesize a type of electrocatalysts made of PGMs anchored on high-surface-area and corrosion-resistive matrix for high performance electrolysis. The produced Pt/Ni-Mo electrocatalyst shows a small overpotential of 113 mV at an ultrahigh current density of 2000 mA cm$^{-2}$ in saline-alkaline water, which is the best performance reported so far. The electrocatalyst also shows a long durability in various electrolytes under harsh conditions including a strong alkaline electrolyte or a simulated seawater electrolyte (24 hours), an ultrahigh current density of 2000 mA cm$^{-2}$ (140 hours), as well as temperatures up to 80 ºC. The production of the catalyst can be scaled up and shows a more competitive performance than catalysts in commercial hydrogen generating equipment. Evaluating from its performance and costs, our results show the great potential of this electrocatalyst in sustainable hydrogen production.

**Data availability**

All data are available from the authors upon reasonable request.

**DECLARATION OF INTERESTS**

Patents related to this research have been filed by Tsinghua-Berkeley Shenzhen Institute, Tsinghua University. The University's policy is to share financial rewards from the exploitation of patents with the inventors.

**ACKNOWLEDGEMENTS**

We acknowledge support from the National Natural Science Foundation of China (Nos. 51722206), the Youth 1000-Talent Program of China, Guangdong Innovative and Entrepreneurial Research Team

Program (No. 2017ZT07C341), the Bureau of Industry and Information Technology of Shenzhen for the "2017 Graphene Manufacturing Innovation Center Project" (No. 201901171523). The authors also thank staff at the BL11B beamline in Shanghai Synchrotron Radiation Facility for their technical assistance.

## AUTHOR CONTRIBUTIONS

F.Y., Y.L. and B.L. conceived the idea. F.Y. synthesized the materials and performed XRD, SEM, and XPS characterization and electrochemical tests. Y.L., Q.Y. and Z.Z. took part in the electrochemical measurements and discussion. Z.Z., S.Z., and J.L. conducted XAS experiments and data analysis. Z.L. and W.R. performed the TEM characterization and analysis. B.L. supervised the project and directed the research. F.Y., Y.L. H.M.C. and B.L. interpreted the results. F.Y., Y.L., L.J and B. L. and wrote the manuscript with feedback from the other authors.

## SUPPLEMENTARY INFORMATION

Supplementary Information can be found online.